\documentclass[12pt]{iopart}

\usepackage{graphicx}
\newcommand \beq{\begin{eqnarray}}
\newcommand \eeq{\end{eqnarray}}
\begin{document}

\title[Hadron-quark continuity induced by the axial anomaly in dense QCD]
{Hadron-quark continuity induced by \\ the axial anomaly in dense QCD}

\author{N Yamamoto$^1$, T Hatsuda$^1$, M Tachibana$^2$ and G Baym$^3$}

\address{$^{1}$Department of Physics, University of Tokyo, Japan\\
$^{2}$Department of Physics, Saga University, Saga 840-8502, Japan\\
$^{3}$Department of Physics, University of Illinois, 1110 W. Green St.,
Urbana, Illinois 61801, USA}
\eads{{yamamoto@nt.phys.s.u-tokyo.ac.jp}}
\begin{abstract}

    We investigate the interplay between the chiral and diquark condensates on
the basis of the Ginzburg-Landau potential with QCD symmetry.  We
demonstrate that the axial anomaly drives a new critical point at low
temperature in the QCD phase diagram and leads to a smooth crossover between
the hadronic and color superconducting phases.

\end{abstract}


\section{Introduction}

    Unravelling the phase structure in the intermediate baryon density region
between the hadronic and deconfined phases is one of the key issues in quantum
chromodynamics (QCD).  The phase structure is also relevant to the matter in
the interiors of neutron and possible quark stars and to the dynamics of
moderate energy heavy-ion collisions in the future at GSI.

    Recently, we have shown the existence of a new critical point and an
associated smooth crossover at intermediate density region in a
model-independent Ginzburg-Landau (GL) analysis of the phase structure
\cite{HTYB06,YHTB07}.  The smooth crossover is intimately connected to the
idea of hadron-quark continuity, proposed in \cite{SW}.  In this analysis, the
coupling between the chiral condensate $\Phi \sim \langle \bar{q}q \rangle$
and the diquark condensate $d \sim \langle qq \rangle$ plays a crucial role.

\section{Ginzburg-Landau potential}

    The guiding principle in constructing the GL potential describing chiral
and diquark condensates is the preservation of the QCD symmetry, ${\cal G}
\equiv SU(3)_L \times SU(3)_R \times U(1)_B \times U(1)_A \times SU(3)_C$,
where $U(1)_A$ is explicitly broken down to $Z_6$ by the axial anomaly.  We
define the chiral condensate ($\Phi$) and diquark condensate ($d$) in the
$J^P=0^+$ channel by $\Phi_{ij} \sim - \langle \bar{q}_R^j q_L^i \rangle$ and
$\langle (q_{_L})_b^j C (q_{_L})_c^k \rangle \sim \epsilon_{abc}
\epsilon_{ijk} [d_{_L}^{\dagger}]_{ai}$ (plus $L \leftrightarrow R)$, where
$i,j,k$ and $a,b,c$ are the flavor and color indices, and
$C=i\gamma^2\gamma^0$ is the charge conjugation operator.  The order
parameters transform under ${\cal G}$ as $\Phi \to {\rm e}^{-2i \alpha_{_A}}
V_{L} \Phi V_{R}^{\dagger}, \ d_L \rightarrow e^{2i\alpha_B + 2i \alpha_A} V_L
d_L V_C^{\rm t}$, where the $V_{L,R,C}$ are $SU(3)_{L,R,C}$ rotations and the
$\alpha_{A,B}$ are $U(1)_{A,B}$ rotations.  Given these tranformations, and
assuming that the order parameters are small enough to write a power series,
we can construct the most general GL potential $\Omega(\Phi,d)$ invariant
under ${\cal G}$ except for $U(1)_A$.  For three massless flavors, after
making an ansatz of maximal flavor symmetry ($\Phi={\rm
diag}(\sigma,\sigma,\sigma)$ and $d_L=-d_R={\rm diag}(d,d,d)$), the general GL
potential assumes the simple form up to fourth order in $\Phi$ and $d$ ,

\beq
 \label{eq:nf3-model}
 \Omega_{\rm 3F}
 &=& \left( \frac{a}{2} \sigma^2 - \frac{c}{3} \sigma^3 + \frac{b}{4}\sigma^4
  + \frac{f}{6}\sigma^6 \right)
  + \left( \frac{\alpha}{2} d^2 + \frac{\beta}{4} d^4 \right) - {\gamma} d^2
  \sigma +  {\lambda} d^2 \sigma^2,
\eeq
where all the coefficient terms depend on temperature, $T$, and chemical
potential, $\mu$.  Since $|\beta| \gg |\lambda|$, as can be shown from
microscopic theories such as weak coupling QCD and the Nambu$-$Jona-Lasinio
(NJL) model, we treat the $\lambda$-term as a small perturbation in the
three-flavor case.  On the other hand, $\gamma$ and $c$, which both originate
from the axial anomaly, are not negligible.  Indeed, a positive $c$ is
essential for making the $\eta'$ meson much heavier than the pion; it also
leads to a first-order chiral phase transition at finite $T$ with $\mu=0$.  In
equation (\ref{eq:nf3-model}), the $\gamma$-term acts as an external field for
$\sigma$ and leads to a new critical point as we see below.

    In principle, the system can have four possible phases:  normal (NOR)
($\sigma=d=0$), CSC ($\sigma=0, d \neq 0$), NG ($\sigma \neq 0, d=0$), and
coexistence (COE) ($\sigma \neq 0, d \neq 0$).  We locate the phase boundaries
and the order of the phase transitions within the GL formalism by comparing
the potential minima in these phases.

\section{Phase structure}

\begin{figure}[t]
\begin{center}
\includegraphics[width=2.8in]{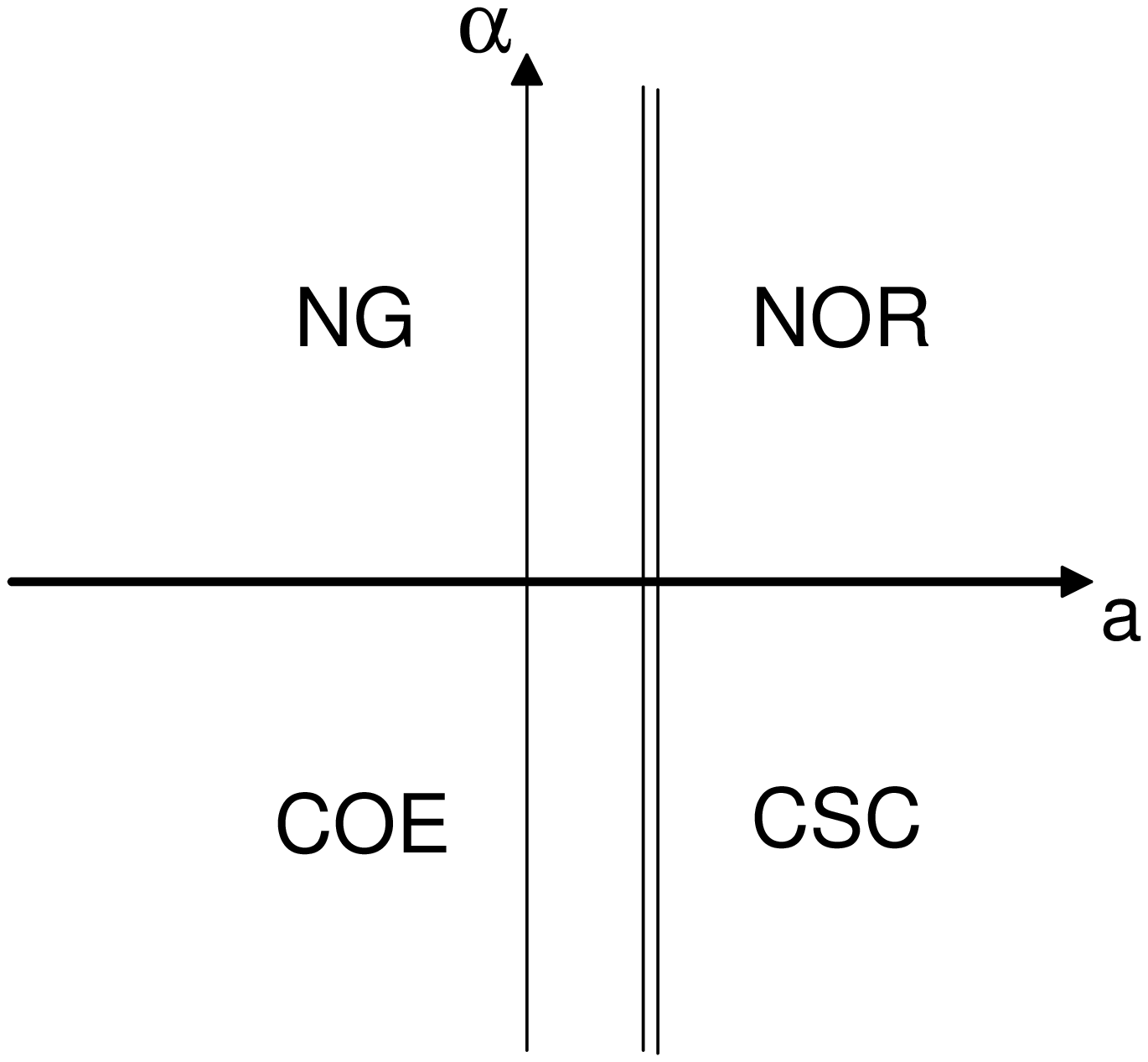}
\hskip 1.5cm
\includegraphics[width=2.6in]{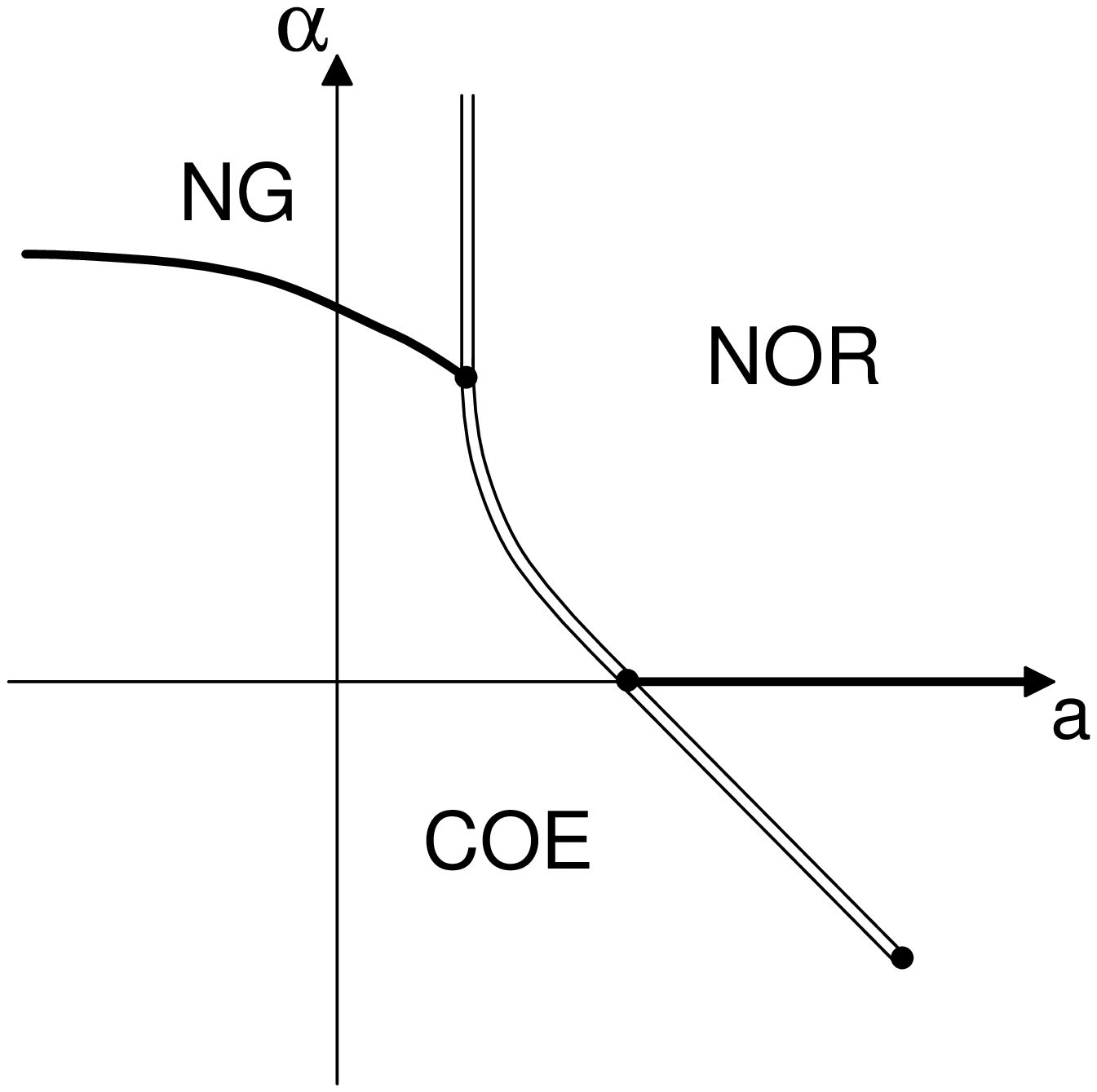}
\caption{Phase structure in the $a$-$\alpha$ plane in the massless
three-flavor case without $\gamma$-coupling (left) and with (right).  Phase
boundaries with a first order transition are denoted by a double line and
second order by a single line.}
\label{fig1}
\end{center}
\vskip -0.3cm
\end{figure}

    Figure~\ref{fig1} shows the phase structure in the $a$-$\alpha$ plane for
the massless three-flavor case with and without the
$\gamma$-coupling.\footnote[2]{For $\gamma > (c/3)\sqrt{\beta/b}$, a
tricritical point emerges on the boundary between the NG and COE phases.} The
figure shows the case for $b >0$; the structure for $b<0$ is not qualitatively
different.  Without the $\gamma$-coupling, the critical lines of the chiral
and the CSC transitions simply cross (left panel of Fig.~\ref{fig1}).  With
the $\gamma$-coupling, the phase structure undergoes major modifications.
Firstly, the first order line between the CSC and COE phases, which originally
continued all the way down terminates at a critical point; as a result, the
CSC and COE phases are continuously connected.  This is because the
$\gamma$-term acts as an external field for $\sigma$, smoothing out the first
order chiral transition for large $\gamma d^2$.  Moreover, the second order
CSC transition splits into two, since $\sigma$ varies discontinuously across
the first order line between the NOR and COE phases.

\begin{figure}[t]
\begin{center}
\includegraphics[width=2.6in]{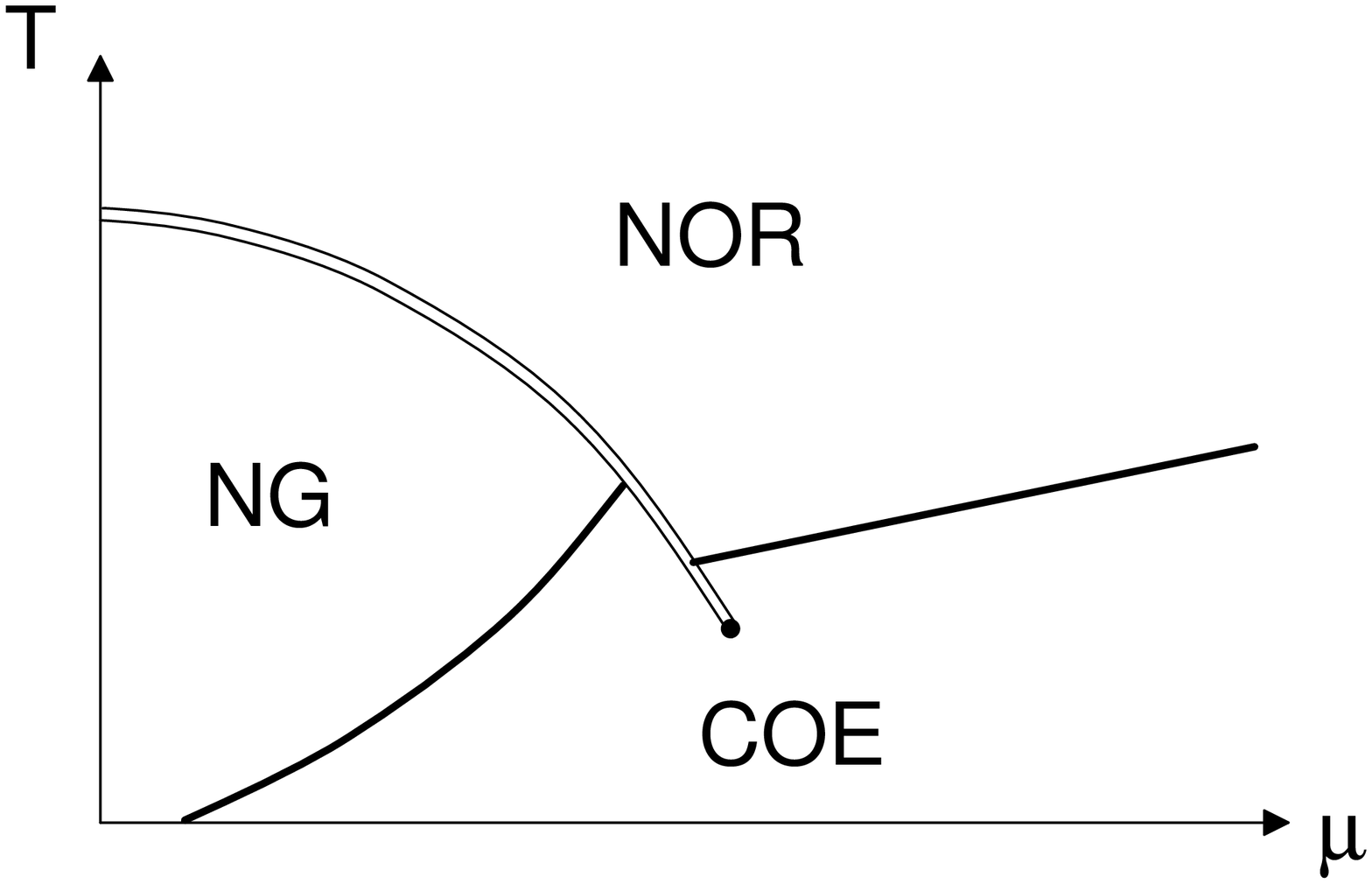}
\hskip 1.5cm
\includegraphics[width=2.6in]{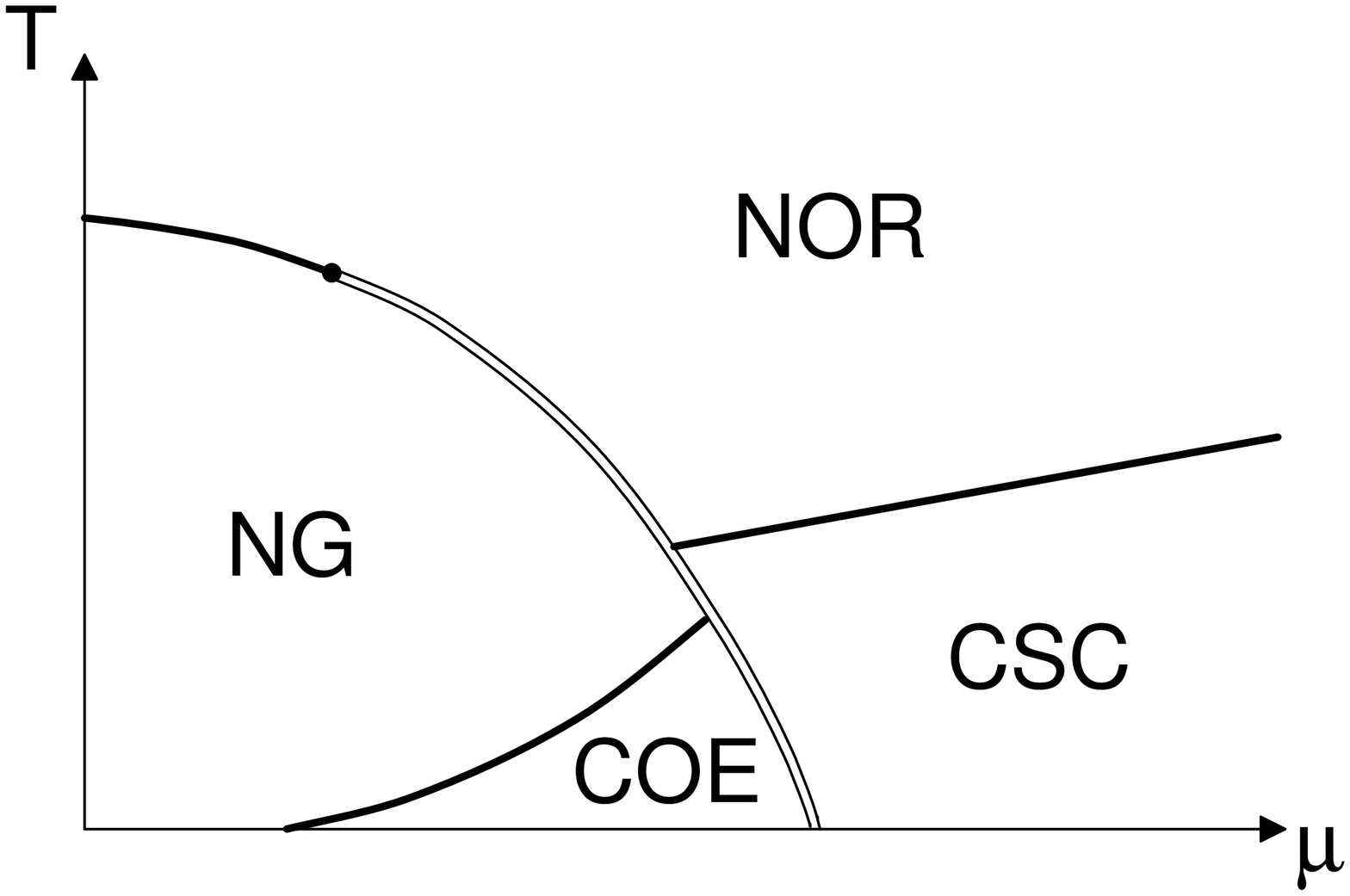}
    \caption{Schematic phase structure in the $T$-$\mu$ plane for the massless
three-flavor case (left) and massless two-flavor case (right).}
\label{fig2}
\end{center}
\vskip -0.3cm
\end{figure}

    Mapping the coordinates from ($a$,$\alpha$) to ($T$,$\mu$), we obtain the
schematic phase diagram for massless three-flavor shown in the left panel of
Fig.~\ref{fig2}.  Analyzing the massless two-flavor case similarly, we arrive
the phase diagram in the right panel of Fig.~\ref{fig2}.

\begin{figure}[t]
\begin{center}
\includegraphics[width=3.0in]{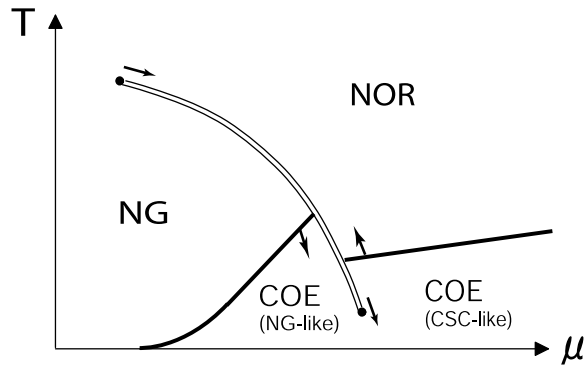}
\caption{Schematic phase structure with realistic quark masses (light up
and down quarks and a medium-heavy strange quark).  The arrows show how the
critical point and the phase boundaries move as the strange-quark mass
increases toward the two-flavor limit.}

\label{fig3}
\end{center}
\vskip -0.3cm
\end{figure}

    The phase structure with realistic quark masses (light up and down quarks
and a medium-heavy strange quark) corresponds to a situation intermediate to
the two cases in Fig.~\ref{fig3}.  There are two critical points in the
diagram.  The first one, near the vertical axis originally found by Asakawa
and Yazaki \cite{AY89}, is driven by the quark mass which washes out the
second order transition in high $T$ region.  The one near the horizontal axis
is our new critical point originating from the axial anomaly.  The existence
of the new critical point indicates that a smooth crossover takes place
between the COE and CSC phases.\footnote{A similar critical point has been
pointed out in \cite{NJL2-model} from the two-flavor NJL model.  However, its
origin is quite different from ours, since the axial anomaly does not produce
the $\sigma d^2$ coupling in two-flavor case.}

\section{Conclusion}

    We have utilized the model-independent GL approach to study the interplay
between the chiral and diquark condensates.  We found that a new critical
point driven by the axial anomaly emerges and leads to a smooth crossover
between the hadronic phase and the color superconducting phase at low
temperature.  Furthermore, in \cite{YHTB07}, we found that there is a smooth
continuity of the elementary excitations, e.g., the ordinary pion ($\bar q q$)
in the NG phase and an exotic pion ($\bar q \bar q qq$) in the CSC phase are
mixed into a generalized pion satisfying a generalized
Gell-Mann$-$Oakes$-$Renner relation in the intermediate density region.  An
interesting problem is to pin down the location of the new critical point from
phenomenological models, and eventually from lattice QCD simulations.

\end{document}